\documentstyle[12pt,graphicx]{article}

\begin{document}

\title{ Masses of Light Mesons \\ and Analytical Confinement}
\author{G. V. Efimov \vspace*{0.2\baselineskip}\\
 \itshape Bogoliubov Laboratory of Theoretical Physics,\\
 \itshape Joint Institute for Nuclear Research, 141980 Dubna,
 Russia\vspace*{0.2\baselineskip} }
%
%
\maketitle
\begin{abstract}
Spectrum of masses of light pseudoscalar ($\pi,~K,~\eta,~\eta'$)
and vector ($\rho,~K^*,~\omega,~\phi$) mesons can be explained on
the base of the following assumptions: (1) analytical confinement
(propagators of quarks and gluons are entire analytical functions
of the Gaussian type), (2) mesons are bound states of quark and
gluons (the Bethe-Salpeter equation in the one-gluon exchange
approximation) and (3) QCD coupling constant $\alpha_s(M)$ is a
monotone decreasing function of mass of a bound state $M$.

The decay constants $f_\pi$ and $f_K$ are calculated.
\end{abstract}

\section{Introduction}

At present time Quantum Chromodynamics (QCD) is considered as the
true theory which control the behavior of quark and gluons as
"bricks" of hadron matter. Due to the confinement phenomena in
experiments we observe colorless hadrons as bound states of quarks
and gluons. Up to now a rigorous analytical solution of the
problem of confinement and hadronization of quarks and gluons into
hadrons is still missing. Therefore effective phenomenological and
semi-phenomenological approaches based on QCD are developed to
derive non-trivial statements about hadronic processes. Because
dynamical mechanism of transformation of quarks and gluons into
hadrons is not clear now the conception of quark and gluon
structures depends on physical context under consideration and the
aim of all theoretical approaches claiming to be obtained from
"the first principles" of QCD is to find process-independent
concepts using symmetry and group arguments (chiral, flavour,
anomaly, mixing and so on) in order to get possibilities to
compare different characteristics of physical processes (there is
huge literature on these problems, see, for example,
\cite{Flamm,Feld}).

We would like to stress that the application of methods of quantum
field theory can be successful if (1) propagators of interacting
particles are known and (2) the effective coupling constant is
small. Namely this situation takes place in the standard model of
electro-weak interactions. What have we in the QCD as theory of
"strong interactions"?

From physical point of view it is evident that processes of
hadronization and confinement of quarks and gluons take place in
the same space-time region, however  the behavior of quarks and
gluons in this confinement region is not known and, from our point
of view, namely this circumstance is the main reason why QFT
methods can not be applied directly. From our point of view if the
behavior of quarks and gluons is known in the confinement region
and the coupling constant is small enough, we will succeed in
analytical solution of hadronization problem.

The idea to get the quark propagator in the confinement region is
not new. The Schwinger-Dyson equation is the main tool to
calculate this propagator (see, for example,
\cite{Roberts,Rob,Prosperi}). However the theoretical problem is
what quark-gluon vertex and what gluon propagator should be used
in this equation and then what computing methods should be applied
to get the solution of this equation. As a result calculations
become to be very cumbersome and opaque (try to repeat these
calculations!).

The confinement is generally accepted to be a result of
nonperturbative nonlinear interactions of gluons in QCD. This idea
plus the Wilson loop confinement argumentation are used to reduce
the relativistic hadronization problem to a stationary
Shr\"{o}dinger picture with an increasing potential. Thus
confinement is a static picture, i.e. there exists a constant in
time potential which keep {\it two} quarks together. Conception of
the confinement of {\it one isolated} quark is not formulated at
all.

Our point of view (see \cite{Ef1}) is that the cause of
confinement is instability of bosons in a homogeneous self-dual
vacuum field. In particular in the quantum electrodynamics
confinement can not take place because all fundamental particles
(leptons and baryons) are fermions. In QCD gluons (bosons!) owing
to nonlinear self-interaction play double role - they transfer the
interaction as photons in QED and play role of real
particles-bosons with zero mass. Namely massless gluons being
bosons leads to instability of QCD-vacuum which is realized for a
nonzero homogeneous self-dual vacuum gluon field which in turn
leads to analytical confinement quark in this vacuum field. Thus
from our point of view the initial "free" quark-gluon Lagrangian
should contain this field and propagators of quarks and gluons
described by this "free" Lagrangian should satisfy the confinement
criterion. Physically analytical confinement means that quarks and
gluons are fluctuations in space and time. Space-time scale of
these fluctuations is defined by the strength of vacuum gluon
field $\Lambda$.

The second question, what is the QCD coupling constant in the
confinement region, i.e. for low energies $E\leq 1~Gev$? Usually
coupling constant is supposed to be constant in this region
although this statement is not completely consistent with general
behavior of the running QCD coupling constant. Various
investigations result in a remarkable variety of its infrared
behavior (see \cite{Schirkov}), so that this question can be
considered to be open. Thus the unique answer this question is not
yet exist, so that some speculations can be done.

The main subject of this paper is the spectrum of masses of light
pseudoscalar ($\pi$, $K$, $\eta$, $\eta'$) and vector ($\rho$,
$K^*$, $\omega$, $\phi$) mesons. The aim of modern theoretical
approaches is to get correlations between masses, i.e. so called
mass formulas on the base of symmetry arguments (see, for example,
\cite{Flamm}). Our point of view if  the propagators of quarks and
gluons are known in the confinement region and the QCD coupling
constant $\alpha_s(M)$ is small enough, then the Bethe-Salpeter
equation can be used to calculate desired masses.

The main result of this paper is that mass differences in
pseudoscalar and vector multiples can be explained on the base of
the following assumptions: (1) analytical confinement (propagators
of quark and gluons are entire analytical functions of the
Gaussian type), (2) mesons are bound states of quark and gluons
(the Bethe-Salpeter equation in the one-gluon exchange
approximation) and (3) QCD coupling constant $\alpha_s(M)$ is a
monotone decreasing function of mass of a bound state $M$.

Our approach is based on the following statements

\begin{itemize}
\item The methods of Quantum Field Theory can be used, it means that
weak coupling constant regime should take place and perturbation
calculations can be applied.

\item Our guess is that the selfdual homogeneous gluon fields with
a constant strength is a good candidate to realize the vacuum QCD.
In this fields {\it analytical confinement} takes place, i.e.
propagators of quarks and gluons are entire analytical functions
in the $p^2$ complex plane.

\item If propagators of constituent particles are known and
coupling constants are small enough  bound states can be found as
solutions of the Bethe-Salpeter equation in the one-gluon exchange
approximation.

\item The variation of the QCD coupling constant $\alpha_s(M)$ in
the low energy region $M\leq 1~Gev$ should be taken into account.
\end{itemize}

Our aim is to understand the general features of spectrum of light
mesons in the most simple dynamical way so that we want to
simplify the problem as much as possible. The first observation is
that the quark and gluon propagators can be approximated by {\it
virton} fields, i.e. by pure Gaussian exponents. Besides solutions
of the Bethe-Salpeter equation in this case can be found in the
explicit analytical form, so that qualitative characteristics of
the mass spectrum can be understood more profoundly (see
\cite{EG}). The second observation is that for light quarks
located in a selfdual homogeneous gluon fields with a constant
strength the main contribution into quark propagators comes from
so called zero modes. We show that these two points defines main
features of light meson spectrum.

Thus our formulation of the problem looks:

\vspace{.5cm}

{\it Does it exist a reasonable form of propagators of quarks and
gluons induced by the behavior of constituents in a selfdual
homogeneous gluon fields with a constant strength and the QCD
coupling constant $\alpha_s(M)$ in the region} $M_\pi\leq M\leq
M_{\eta'}$ {\it for which one can obtain the masses of
pseudoscalar and vector mesons ?}

\vspace{.5cm}

\section{Lagrangian and propagators.}

Our basic assumption is that the QCD vacuum is realized by a
self-dual gluon field with constant strength
\begin{eqnarray}
\label{B} && \breve{B}_\mu(x)=B_\mu^a(x)t^a,~~~~~
B_\mu^a(x)=n^aB_\mu(x),~~~~~ B_\mu(x)=\Lambda^2~b_{\mu\nu}x_\nu,\\
&& B_{\mu\nu}(x)=\partial_\mu B_\nu(x)-\partial_\nu B_\mu(x)=
-2\Lambda^2~b_{\mu\nu}={\rm const}.\nonumber
\end{eqnarray}
Here $n^a$ is a constant vector in color space. The parameter
$\Lambda$ defines the confinement scale and
\begin{eqnarray*}
&& b_{\mu\nu}=-b_{\nu\mu},~~~~b_{\mu\rho}b_{\rho\nu}=
-\delta_{\mu\nu},~~~~\tilde{b}_{\mu\nu}=
{1\over2}\epsilon_{\nu\mu\alpha\beta}b_{\alpha\beta}=\pm
b_{\mu\nu}.
\end{eqnarray*}
The field $\breve{B}_\mu(x)$ satisfies the Yang-Mills equations.
The standard QCD Lagrangian in this field looks like
\begin{eqnarray}
\label{LQCD} &&{\cal L}=-{1\over8}{\rm Tr}~{\breve G}_{\mu\nu}^2
+\sum\limits_f\left(\bar{q}_f\left[\gamma_\mu(\breve{\nabla}_\mu+
ig\breve{A}_\mu)-m_f\right]q_f\right)\\ && {\breve
G}_{\mu\nu}(x)=\breve{\nabla}_\nu{\breve A}_\mu-
\breve{\nabla}_\mu{\breve A}_\nu+g[{\breve A}_\mu(x),{\breve
A}_\nu(x)], ~~~~~{\breve A}_\mu=t^aA_\mu^a\nonumber
\end{eqnarray}
where
\begin{eqnarray*}
&& \breve{\nabla}_\mu(x)=\partial_\mu+i\breve{B}_\mu(x)
=\partial_\mu+i\Lambda^2\breve{n}b_{\mu\nu}x_\nu,\\ &&
[\breve{\nabla}_\mu(x),\breve{\nabla}_\nu(x)]=
-2i\Lambda^2\breve{n}b_{\mu\nu},~~~~~~\breve{n}=n^at^a.\nonumber
\end{eqnarray*}

The part of the QCD Lagrangian which is responsible for meson
hadronization can be written in the form
\begin{eqnarray}
\label{lagr} && {\cal L}=(\bar{q}S^{-1}q)-{1\over2}(gD^{-1}g)+
g(\bar{q}i\gamma_\mu t^a q)g_\mu^a,
\end{eqnarray}
where the quark field $q(x)=q_{fa\alpha}(x)$ has indexes
$$f~-~{\rm flavor}~SU_f(3),~~ a~-~{\rm
color}~SU_c(3),~~\alpha~-~{\rm spin}$$ and the gluon field
$g_{a\mu}(x)$ has $$a~-~{\rm color}~SU_c(3),~~\mu~-~{\rm
vector}.$$
 The quark flavor spinor can be represented as
\begin{eqnarray*}
&& SU_f(3):~~~~~~~~q=\left(\begin{array}{c}
u \\
d\\
s\\
\end{array}\right)
\end{eqnarray*}

Quark propagator in the self-dual homogeneous gluon vacuum field
is the solution of the equation (see details in \cite{Efned})
\begin{eqnarray*}
&&(\hat{\nabla}(x)-m)S(x)=-\delta(x),~~~~~
\nabla_\mu(x)=\partial_\mu+i\Lambda^2\breve{n}b_{\mu\nu}x_\nu.
\end{eqnarray*}
It has the form
\begin{eqnarray}
\label{Propq} \tilde{S}_\pm(p)&=&\int\limits_0^1
{du\over2\Lambda^2} e^{-u{p^2\over2\Lambda^2}}
\left({1-u\over1+u}\right)^{{m^2\over4\Lambda^2}}\\
&\cdot&\left\{i\hat{p}\pm u\breve{n}\gamma_5(\gamma bp)
+{m\over1-u^2} \left[1\mp\gamma_5u^2+{i\over2}\breve{n}(\gamma
b\gamma)u\right]\right\}. \nonumber
\end{eqnarray}
The gluon propagator is obtained in \cite{BE}. We do not show it
here because formula is quite cumbersome.

 We shall use the rough approximation of propagators
preserving the main their features. The quark propagator having
the Gaussian form and "zero mode" behavior for small quark mass
$\sim{1\over m}$ is chosen in the form
\begin{eqnarray}
\label{Pqa}
 \tilde{S}_f(p)&=&\left({i\hat{p}\over\Lambda^2}+
{1\pm\omega_f\gamma_5\over m_f}\right)e^{-{p^2\over2\Lambda^2}}=
\tilde{s}_f(p)e^{-{p^2\over2\Lambda^2}},\\
\tilde{s}_f(p)&=&{i\hat{p}\over\Lambda^2}+
{1\pm\omega_f\gamma_5\over m_f},~~~~~~
\omega={1\over1+{m^2_f\over\Lambda^2}}.\nonumber
\end{eqnarray}

The gluon propagator is chosen in the form
\begin{eqnarray*}
&& D^{aa'}_{\mu\mu'}(x-x')=\langle
g_{a\mu}(x)g_{a'\mu'}(x')\rangle=
\delta^{aa'}\delta_{\mu\mu'}D(x-x')
\end{eqnarray*}
where
\begin{eqnarray}
\label{Pga}
 &&\tilde{D}(k)={1\over\Lambda^2}e^{-{k^2\over2\Lambda^2}},~~~~~~
 D(y)={\Lambda^2\over(2\pi)^2}e^{-{\Lambda^2y^2\over2}}
\end{eqnarray}

This rough choice of the quark and gluon propagators (\ref{Pqa})
and (\ref{Pga}) is defined by the unique reason only: the
Bethe-Salpeter equation can be solved analytically in this case
and we get simple analytical formulas for the meson spectrum (see
also \cite{EG}).

\section{One-gluon exchange}

In one-gluon exchange the four-quark interaction Lagrangian is
\begin{eqnarray}
\label{W2}
 W&=&{g^2\over2}\int\!\!\!\int
dx_1dx_2(\bar{q}(x_1)i\gamma_\mu t^a q(x_1))
D(x_1-x_2)(\bar{q}(x_2)i\gamma_\mu t^a q(x_2))
\end{eqnarray}

In order to go to colorless quark currents the Fierz
transformations should be done
\begin{itemize}
\item Color transformations
\begin{eqnarray*}
&& SU_c(3):~~~~~~ (t^a)_{j_1j_1'}(t^a)_{j_2j_2'}=
{4\over9}\delta_{j_1j_2'}\delta_{j_2j_1'}-
{1\over3}(t^a)_{j_1f_2'}(t^a)_{f_2f_1'}
\end{eqnarray*}

\item Flavor transformations
\begin{eqnarray*}
&& SU_f(3):~~~~~~ \delta_{f_1f_1'}\delta_{f_2f_2'}=
{1\over3}\delta_{f_1f_2'}\delta_{f_2f_1'}+
{1\over2}(\lambda^a)_{f_1f_2'}(\lambda^a)_{f_2f_1'}
\end{eqnarray*}

\item Dirac spin transformations
\begin{eqnarray*}
(i\gamma_\mu)_{\alpha_1\alpha_1'}(i\gamma_\mu)_{\alpha_2\alpha_2'}&=&
-{1\over2}(i\gamma_\mu)_{\alpha_1\alpha_2'}(i\gamma_\mu)_{\alpha_2\alpha_1'}
-(i\gamma_5)_{\alpha_1\alpha_2'}(i\gamma_5)_{\alpha_2\alpha_1'}\\
&-&I_{\alpha_1\alpha_2'}I_{\alpha_2\alpha_1'}+
{1\over2}(\gamma_\mu\gamma_5)_{\alpha_1\alpha_2'}
(\gamma_\mu\gamma_5)_{\alpha_2\alpha_1'}
\end{eqnarray*}
\end{itemize}

Then for the four-quark interaction Lagrangian with pseudoscalar
and vector colorless quark currents we get
\begin{eqnarray}
\label{WP}
 W_P&=&{1\over2}\cdot{4g^2\over9}\int\!\!\!\int
dx_1dx_2~D(x_1-x_2)\\
&\cdot&\left\{{1\over3}(\bar{q}(x_1)i\gamma_5q(x_2))
(\bar{q}(x_2)i\gamma_5q(x_1))+
{1\over2}(\bar{q}(x_1)i\gamma_5\lambda^a q(x_2))
(\bar{q}(x_2)i\gamma_5\lambda^a q(x_1))\right\},\nonumber
\end{eqnarray}
\begin{eqnarray}
\label{WV}
 W_V&=&{1\over2}\cdot{2g^2\over9}\int\!\!\!\int
dx_1dx_2~D(x_1-x_2)\\
&\cdot&\left\{{1\over3}(\bar{q}(x_1)i\gamma_\mu q(x_2))
(\bar{q}(x_2)i\gamma_\mu q(x_1))+
{1\over2}(\bar{q}(x_1)i\gamma_5\lambda^a q(x_2))
(\bar{q}(x_2)i\gamma_5\lambda^a q(x_1))\right\}.\nonumber
\end{eqnarray}

Flavor structure of meson currents looks like
\begin{eqnarray*}
&&{1\over3}(\bar{q}q)(\bar{q}q)+{1\over2}(\bar{q}\lambda^a q)
(\bar{q}\lambda^a q)\\
&&=\left[\cos\theta(\bar{n}n)+\sin\theta(\bar{s}s)\right]^2+
\left[-\sin\theta(\bar{n}n)+\cos\theta(\bar{s}s)\right]^2\\
&&+(\bar{n}\tau^j n) (\bar{n}\tau^j n)+
2[(\bar{u}s)(\bar{s}u)+(\bar{d}s)(\bar{s}d)]
\end{eqnarray*}
where
 $$ n={1\over2^{1/4}}\left(\begin{array}{c} u \\ d\\
\end{array}\right),
~~~~~~(\bar{n}n)={(\bar{u}u)+(\bar{d}d)\over\sqrt{2}}$$

The ideal mixing angle is
$$\theta_{id}=\arcsin{1\over\sqrt{3}}=35.3^\circ$$

Pseudoscalar nonet is defined as
\begin{eqnarray}
\label{PP}
 && \pi^j=\bar{n}\tau^jn,~~~~~~K^-=\bar{u}s,\\ &&
\eta=\cos\theta\cdot{\bar{u}u+\bar{d}d\over\sqrt{2}}+
\sin\theta\cdot\bar{s}s,~~~~~~~
\eta'=-\sin\theta\cdot{\bar{u}u+\bar{d}d\over\sqrt{2}}+
\cos\theta\cdot\bar{s}s.\nonumber
\end{eqnarray}

Vector nonet is defined as
\begin{eqnarray}
\label{VV}
&&\rho^j=\bar{n}\tau^jn,~~~~~~K^*=\bar{u}s,~~~~~~\omega=\bar{n}n,
~~~~~~~\phi=\bar{s}s.
\end{eqnarray}

\section{Solution of the Bethe-Salpeter equation}

The solution of the Bethe-Salpeter equation is reduced to the
following variation problem (see \cite{BS})
\begin{eqnarray}
\label{lambda}
 && \lambda_J(M)=-{4g^2\over9}
\max\limits_U{(U\sqrt{D}~\Pi_J\sqrt{D}U) \over (UU)},~~~~J=P,~V.
\end{eqnarray}
Here
\begin{eqnarray*}
\Pi_J(p|y_1+y_2)&=&\int dx e^{i(px)}C_J{\rm Tr}
\left[\Gamma_JS_1\left(x-{y_1+y_2\over2}\right)
\Gamma_JS_2\left(-x-{y_1+y_2\over2}\right)\right]\\
&=&\int{dk\over(2\pi)^4}e^{i(k(y_1+y_2))}\tilde{\Pi}_J(k,p)\end{eqnarray*}
$$ C_P=1,~~~~~~C_V={1\over2}$$
\begin{eqnarray*}
\tilde{\Pi}_P(k,p)&=&{\rm Tr}\left[i\gamma_5\tilde{S}_1
\left(k+{p\over2}\right)i\gamma_5\tilde{S}_2
\left(k-{p\over2}\right)\right]\\
&=&-e^{-{k^2\over\Lambda^2}+{M^2\over4\Lambda^2}}
{48\over\Lambda^2}\left[{k^2\over2\Lambda^2}+{M^2\over8\Lambda^2}+
{\Lambda^2\over m_1m_2}{1+\omega_1\omega_2\over2}\right]
\end{eqnarray*}

\begin{eqnarray*}
\tilde{\Pi}_V(k,p)&=&{1\over2}{\rm Tr}\left[i\gamma_\mu\tilde{S}_1
\left(k+{p\over2}\right)i\gamma_\nu\tilde{S}_2
\left(k-{p\over2}\right)\right]\\
&=&-\delta_{\mu\nu}e^{-{k^2\over\Lambda^2}+{M^2\over4\Lambda^2}}
{12\over\Lambda^2}\cdot\left[{k^2\over2\Lambda^2}+{M^2\over4\Lambda^2}+
{\Lambda^2\over m_1m_2}\cdot(1-\omega_1\omega_2)\right]
\end{eqnarray*}

$$ {\rm Tr}:~~~4\cdot3\cdot2={\rm spin}\cdot{\rm color}\cdot{\rm
duality}$$

To solve the variation problem (\ref{lambda}) we choose the test
function in the form
\begin{eqnarray}
\label{Utest}
&& U(y)=e^{-a{\Lambda^2y^2\over4}},~~~~~(U U)=\int
dy e^{-a{\Lambda^2y^2\over2}}={(2\pi)^2\over a^2\Lambda^4}
\end{eqnarray}
Then we get
\begin{eqnarray*}
&& U(y)\sqrt{D(y)}={\Lambda\over2\pi}
e^{-(a+1){\Lambda^2y^2\over4}}
\end{eqnarray*}
Let us define the vertex function
\begin{eqnarray}
\label{Vertex} && V(y)={U(y)\sqrt{D(y)}\over\sqrt{(UU)}}=
{\Lambda^3~a\over(2\pi)^2}e^{-(a+1){\Lambda^2y^2\over4}},\nonumber\\
&& \tilde{V}(k)=\int dy V(y)e^{i(ky)}={4~a\over\Lambda(a+1)^2}
e^{-{k^2\over\Lambda^2}\cdot{1\over a+1}}.
\end{eqnarray}
The variation problem (\ref{lambda}) is reduced to
\begin{eqnarray}
\label{lvar}
&&\lambda_J(M)=-{4g^2\over9}\max\limits_a(V\Pi_JV),~~~~J=P,~V.
\end{eqnarray}
 Let us consider
\begin{eqnarray*}
&&(V\Pi_JV)=\int{dk\over(2\pi)^4}\tilde{V}^2(k)\tilde{\Pi}_J(k,p)
\end{eqnarray*}
with
\begin{eqnarray*}
&&\int{dk\over(2\pi)^4}e^{-{k^2\over\Lambda^2}-
{k^2\over\Lambda^2}\cdot{2\over a+1}}
\left[{k^2\over2\Lambda^2}+A\right]
={\Lambda^4\over(4\pi)^2}\cdot{(1+a)^2\over(3+a)^2}\cdot
\left[{1+a\over 3+a}+A\right]
\end{eqnarray*}
We get
\begin{eqnarray}
\label{FA}
F(A)&=&{4\cdot16\cdot48\over9\cdot4}
\max\limits_a\left({a\over(1+a)(3+a)}\right)^2
\left[{1+a\over 3+a}+A\right]\\
&=&{256\over3}\left({a_{max}(A)\over(1+a_{max}(A))(3+a_{max}(A))}\right)^2
\left[{1+a_{max}(A)\over 3+a_{max}(A)}+A\right]\approx F_0+F_1\cdot A
\nonumber
\end{eqnarray}
where
 $$ F_0=0.9188,~~~~~~~F_1=1.52~. $$
The point $a_{max}(A)$ is defined by one of three roots of
variation equation which is the algebraic equation of the third
oder
\begin{eqnarray*}
&&\max\limits_a\left({a\over(1+a)(3+a)}\right)^2\left[{1+a\over
3+a}+ A\right],\\ && -3-4a+a^3+A(-9-3a+3a^2+a^3)=0
\end{eqnarray*}
and can be found in the explicit form
\begin{eqnarray}
\label{aA}
 && a_{max}(A)\approx1.7238+{0.3268\over A+0.5646}
\end{eqnarray}

Thus the eigenvalues of the Bethe-Salpeter equation can be written
in the explicit analytical form.

For pseudoscalar mesons one can get
\begin{eqnarray}
\label{lPP}
 -\lambda_P(M_P)
&=&{\alpha_s\left({M_P^2\over\Lambda^2}\right)\over\pi}
e^{{M_P^2\over4\Lambda^2}}
F\left({M_P^2\over8\Lambda^2}+H_P\right)
\end{eqnarray}
where
\begin{eqnarray*}
&& H_P=\left\{\begin{array}{l} H_\pi={\Lambda^2\over m_u^2}
\cdot{1+\omega_u^2\over2}\\
\\
H_K={\Lambda^2\over m_u m_s}\cdot{1+\omega_u\omega_s\over2}\\
\\
H_\eta={\Lambda^2\over
m_u^2}\cdot{1+\omega_u^2\over2}\cos^2\theta+ {\Lambda^2\over
m_s^2}\cdot{1+\omega_s^2\over2}\sin^2\theta\\
\\
H_{\eta'}={\Lambda^2\over
m_u^2}\cdot{1+\omega_u^2\over2}\sin^2\theta+{\Lambda^2\over
m_s^2}\cdot{1+\omega_s^2\over2}\cos^2\theta\\
\end{array}\right.
\end{eqnarray*}

For vector mesons we get
\begin{eqnarray}
\label{lVV}
 -\lambda_V(M_V)
&=&{\alpha_s\left({M_V^2\over\Lambda^2}\right)\over\pi}
e^{{M_V^2\over4\Lambda^2}}{1\over 4}
F\left({M^2\over4\Lambda^2}+H_V\right)
\end{eqnarray}
where
\begin{eqnarray*}
&& H_V=\left\{\begin{array}{l} H_\rho={\Lambda^2\over m_u^2}
(1-\omega_u^2)\\
\\
H_{K^*}={\Lambda^2\over m_u m_s}\cdot(1-\omega_u\omega_s)\\
\\
H_\omega={\Lambda^2\over m_u^2}\cdot(1-\omega_u^2)\\
\\
H_{\phi}={\Lambda^2\over m_s^2}\cdot(1-\omega_s^2)\\
\end{array}\right.
\end{eqnarray*}

\section{Masses of pseudoscalar and vector mesons}

The experimental masses  of pseudoscalar and vector mesons are (in
$Mev$)
\begin{eqnarray}
\label{masses}
&& M_\pi=140,~~~~ M_K=495,~~~~ M_\eta=547,~~~~~
M_{\eta'}=958,\\ && M_\rho=770,~~~~ M_\omega=782,~~~~
M_{K^*}=892,~~~~~ M_\phi=1019. \nonumber
\end{eqnarray}
We have four free parameters
\begin{eqnarray}
\label{Param}
 && \Lambda,~~~~~m_u,~~~~~m_s,~~~~~\theta
\end{eqnarray}
where $\Lambda$ characterizes the confinement scale, $m_u=m_d$ and
$m_s$ are quark masses, $\theta$ is the mixing angle.

The eigenvalues of the Bethe-Salpeter equation on the masses of
our mesons look like
\begin{eqnarray}
\label{lPV} -\lambda_P(v_P)&=&
{\alpha_s(v_P)\over\pi}e^{{v_P\over4}}
F\left({v_P\over8}+H_P\right),~~~~~
v_P=\left({M_P\over\Lambda}\right)^2
\end{eqnarray}
\begin{eqnarray*}
-\lambda_V(v_V)&=& {\alpha_s(v_V)\over\pi}e^{{v_V\over4}}{1\over4}
F\left({v_V\over4}+H_V\right),~~~~~
v_V=\left({M_V\over\Lambda}\right)^2
\end{eqnarray*}
and these eigenvalues should satisfy the equations
\begin{eqnarray}
\label{Eq}
 && 1+\lambda(v_P)=0,~~~~~~1+\lambda(v_V)=0
\end{eqnarray}

We suppose that there exists the coupling constant $\alpha_s(v)$,
where $\alpha_s(v)$ is a monotone decreasing function and in the
points $v_P$ and $v_V$ this function satisfies
\begin{eqnarray*}
&& \alpha_s(v_P)=\pi R_P(v_P),~~~~~
\alpha_s(v_V)=\pi R_V(v_V)
\end{eqnarray*}
where
\begin{eqnarray}
\label{RPV}
 R_P(v_P)&=&{e^{-{v_P\over4}}\over F\left({v_P\over8}+
H_P\right)},~~~~~P=\{\pi,~K,~\eta,~\eta'\},\\
R_V(v_V)&=&{4~e^{-{v_V\over4}}\over
F\left({v_V\over4}+H_V\right)},
 ~~~~~V=\{\rho,~K^*,~\omega,~\phi\}.\nonumber
\end{eqnarray}

We consider that the experimental masses  of pseudoscalar and
vector mesons are fixed by (\ref{masses}). The problem is to find
the parameters $\Lambda,~m_u,~m_s,~\theta$ in such a way that a
monotone decreasing function $R(v)$ should be {\it smooth as much
as possible} and satisfy (\ref{RPV}). In our examples we have
selected these parameters by eye and then have used the
"Mathematica" program
\begin{eqnarray}
\label{Rv}
 R(v)&=&{\rm Fit}[\{v_P,~R_P(v_P)\},~\{v_V,~R_V(v_V)\},
~\{1,v,v^2,...,v^n\},~v]
\end{eqnarray}
for some $n$.

Thus the coupling constant can be calculated
\begin{eqnarray}
\label{asv}
 && \alpha_s(v)=\pi R(v)
\end{eqnarray}

\section{Effective coupling constant and meson currents.}

The eigenvalues of the Bethe-Salpeter equation are the
polarization operators of corresponding mesons.  Let us consider
the pseudoscalar mesons. We have
\begin{eqnarray}
\label{proc} &&
\Pi_P(v)=\lambda_P(v)=R(v_P)e^{{v\over4}}F\left({v\over8}+H_P\right)
\end{eqnarray}

The kinetic term in the effective Lagrangian of pseudoscalar
mesons looks in a vicinity of the mass shell
\begin{eqnarray*}
&& \left(P[1-\lambda_P(v)]P\right)\rightarrow
\left(P[1-\lambda_P(v_P)-\lambda_P'(v_P)(v-v_P)]P\right)\\
&&=\left(P[Z_P(v_P-v)]P\right)=
\left(P\left[{Z_P\over\Lambda^2}(M_P^2-p^2)\right]P\right)
\rightarrow(P[M_P^2-p^2]P),
\end{eqnarray*}
where the renormalization of the meson field is done
\begin{eqnarray*}
&& P(x)\rightarrow {\Lambda\over\sqrt{Z_P}}P(x)
\end{eqnarray*}
Here the constant of renormalization is
\begin{eqnarray}
\label{ZP}
 && Z_P=\left.{\partial\over\partial
v}\lambda_P(v)\right|_{v=v_P}=
\lambda_P'(v_P)={1\over4}+{F'(v_P)\over F(v_P)}
\end{eqnarray}

The renormalization of the meson fields leads to the
renormalization of the coupling constant and therefore to the
renormalization of the vertexes which determine the binding of the
meson fields with the quark currents.

For the vertex the renormalization leads to
\begin{eqnarray}
\label{gVP}
 g\tilde{V}_P&\rightarrow&
g{\Lambda\over\sqrt{Z_P}}{1\over\Lambda}{\cal V}_P(k)=
{g\over\sqrt{Z_P}}{\cal V}_P(k)=g_P{\cal V}_P(k)\nonumber\\
&=&2\pi\sqrt{{\alpha_s(v_P)\over\pi Z_P}}\cdot
{4a_P\over(a_P+1)^2} e^{-{k^2\over\Lambda^2}\cdot{1\over a_P+1}}
\end{eqnarray}
where
\begin{eqnarray*}
&& a_P=a_{max}\left({v_P\over8}+H_P\right).
\end{eqnarray*}

The interaction Lagrangian of the pseudoscalar meson $P$ with
quarks is
\begin{eqnarray}
\label{LI} && L_I(x)={2g_P\over3}\left(\bar{q}(x){\cal V}_P
\left({1\over2}\stackrel{\leftrightarrow}{p}\right)\lambda_P
i\gamma_5 q(x)\right)P(x)
\end{eqnarray}
with an appropriate matrix $\lambda_P$.

\section{The decay constants $f_P$}

The decay constants $f_P$ are defined by the matrix elements
\begin{eqnarray}
\label{TP}
&& T_{P\mu}^a(p)=\langle0|J^a_{\mu 5}(0)|P(M_P)\rangle=
if^a_P~p_\mu.
\end{eqnarray}
The axial vector local currents with quantum numbers of pion and
kaon are
\begin{eqnarray*}
&& J^\pi_{\mu 5}(x)={1\over\sqrt{2}}\left(\bar{u}(x)
\gamma_\mu\gamma_5 d(x)\right),~~~~~~ J^K_{\mu
5}(x)={1\over\sqrt{2}}\left(\bar{u}(x)\gamma_\mu \gamma_5
s(x)\right).
\end{eqnarray*}
The matrix element (\ref{TP}) is defined by the integral
\begin{eqnarray*}
T_{P\mu}^{12}(p)&=&{2g_P\over3\sqrt{2}}\int{dk\over(2\pi)^4}{\cal
V}_P(k) {\rm Tr}\left[i\gamma_5\tilde{S}_1\left(k+{p\over2}\right)
\gamma_\mu\gamma_5\tilde{S}_2\left(k-{p\over2}\right)\right]
\end{eqnarray*}
After simple calculations one can get
\begin{eqnarray}
\label{fPK} f_\pi&=&
{2\Lambda\over3\sqrt{2}}\cdot\sqrt{{\alpha_s(v_\pi)\over\pi
Z_\pi}} \cdot{12a_\pi\over\pi(2+a_\pi)^2}\cdot e^{{v_P\over4}}
\cdot {\Lambda\over m_u},\\ f_K&=&
{2\Lambda\over3\sqrt{2}}\cdot\sqrt{{\alpha_s(v_K)\over\pi Z_K}}
\cdot{12a_K\over\pi(2+a_K)^2}\cdot e^{{v_K\over4}} \cdot {1\over2}
\left[ {\Lambda\over m_u}+{\Lambda\over m_s}\right].\nonumber
\end{eqnarray}

\section{Results and conclusion}

The numerical results of our calculations are shown on the Figures
1-3. They should be considered as preliminary qualitative
estimations. The main conclusions are
\begin{itemize}
\item Spectrum of meson masses can be understood as result of
(1) the analytical confinement and (2) a monotone decreasing
coupling constant.
\item Decay constants $f_\pi$ and $f_K$ have reasonable values.
\item Different character of the behavior of $\alpha_s(v)$ on
Figures 1-3 means that mass spectrum of pseudoscalar and vector
mesons can not fix parameters (\ref{Param}) uniquely. We should
take into account other matrix elements ($\pi \to\gamma\gamma$,
$\rho\to\pi\pi$ and so on).
\end{itemize}

Thus the simple form of quark and gluon propagators gives
qualitative correct results, so that we can say that the structure
of the quark-gluon interaction in the confinement region is
guessed correctly. Besides we want to stress that  relativistic
scheme of hadonization was used and this scheme is very far from
nonrelativistic potential picture.

Another point is to connect the behavior of the coupling constant
$\alpha_s\left({M^2\over\Lambda^2}\right)$ in the region
$M\leq1000~Mev$ with known behavior in the region $M\geq 1000~Mev$
to compare with available results \cite{Schirkov}.

This work was supported by the grant RFBR 04-02-17370.

\begin{figure}[ht]
\vspace{-.3cm} \centerline{
\includegraphics[width=0.9\textwidth,angle=270]{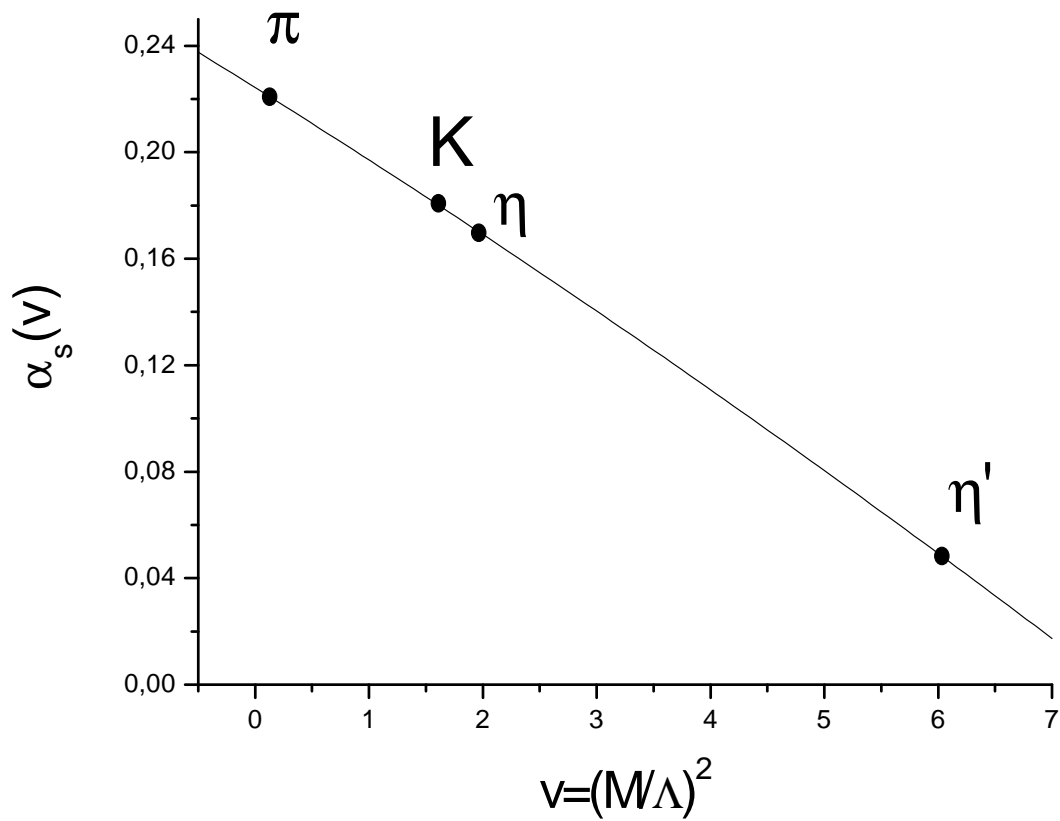}
} \caption{Pseudoscalar mesons. Parameters
$\Lambda=390,~m_u=200,~~m_s=260,~~\theta=54^\circ.$ Decay
constants $f_\pi=133,~~f_K=155$. }
 \label{Fig1} \vspace*{-.3cm}
\end{figure}

\begin{figure}[ht]
\vspace{-.3cm} \centerline{
\includegraphics[width=0.9\textwidth,angle=270]{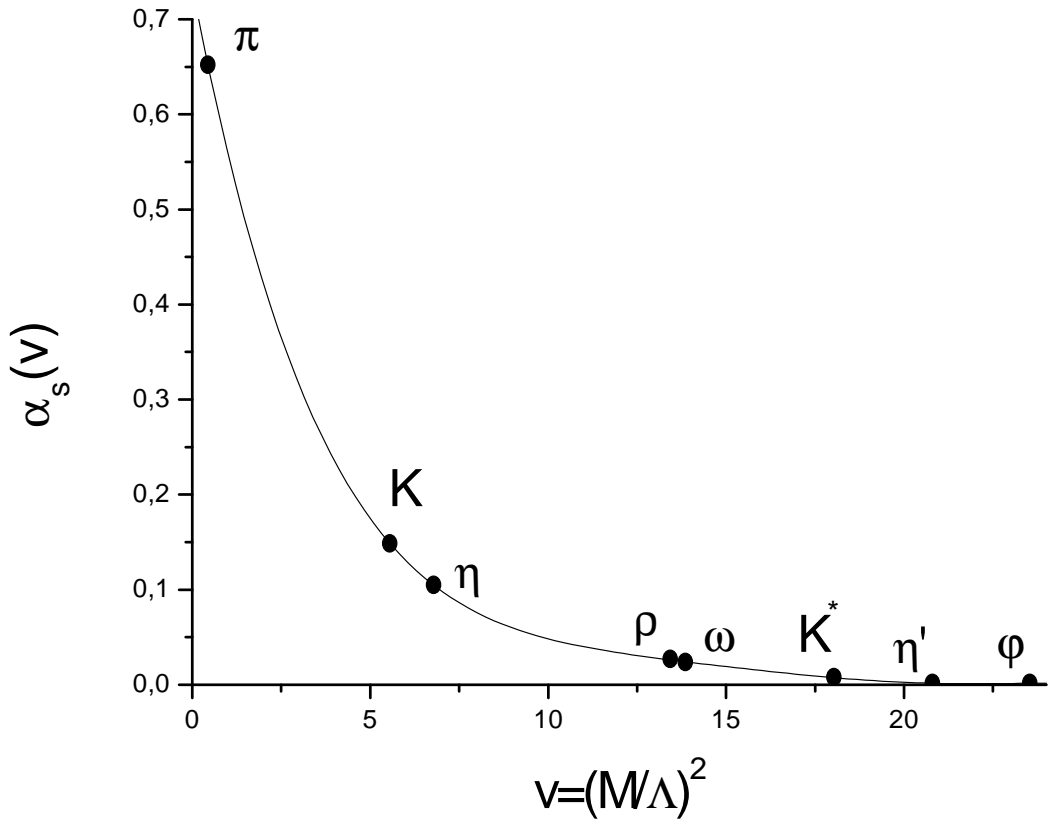}
} \caption{ Pseudoscalar and vector mesons. Parameters
$\Lambda=210,~m_u=180,~~m_s=230,~~\theta=54^\circ.$ Decay
constants $f_\pi=80,~~f_K=107$. }\label{Fig2} \vspace*{-.3cm}
\end{figure}

\begin{figure}[ht]
\vspace{-.3cm} \centerline{
\includegraphics[width=0.9\textwidth,angle=270]{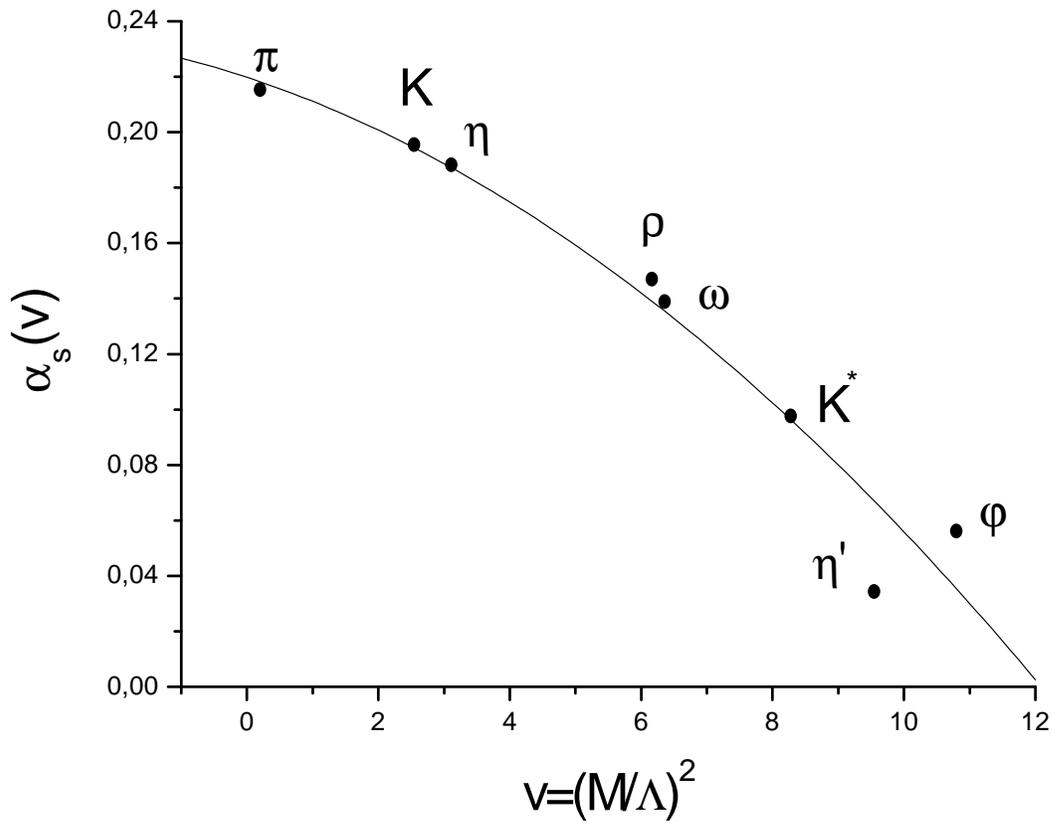}
} \caption{ Pseudoscalar and vector mesons.  Parameters
$\Lambda=310,~m_u=130,~~m_s=240,~~\theta=64^\circ.$ Decay
constants $f_\pi=131,~~f_K=172$.} \label{Fig3} \vspace*{-.3cm}
\end{figure}

\end{document}